\documentclass[preprint]{aastex}

\shorttitle{Edge-on galaxies from 2MASS images}
\shortauthors{Mitronova \& Bizyaev}

\begin{document}

\title{Structural Parameters of Stellar Disks in Edge-on Galaxies \\
from 2MASS Images
}

\author{Mitronova S.\altaffilmark{1} and Bizyaev D. \altaffilmark{2,3}}

\altaffiltext{1}{Special Astrophysical Observatory, Nizhniy Arkhyz, Russia}

\altaffiltext{2}{National Optical Astronomy Observatory, Tucson, AZ, USA}

\altaffiltext{3}{Sternberg Astronomical Institute, Moscow, Russia}

\begin{abstract}

We analyze the J, H, and Ks 2MASS images of 140 late-type edge-on galaxies
selected from the RFGC catalog (which contains flat galaxies with
major-to-minor axis ratio a/b $>$ 7). The NIR scalelengths (h) and
scaleheights (z$_0$) of the stellar disks are determined for all selected
galaxies. The mean relative ratios of the scaleheights of their stellar
disks are 1.00:0.91:0.86 in J:H:Ks bands, respectively.  We infer that
the scaleheight determined from the Ks-band images is, on average, 13\%
larger than the extinction-free scaleheight. This difference is much larger
if the scaleheights were found from the optical-band images.  The relative
thickness (z$_0$/h) of the stellar disks correlates well with their deprojected
central surface brightness obtained from the 2MASS images. This project was
partially supported by grant RFBR 04-02-16518.

\end{abstract}

\section{Introduction, Data, and Analysis}

Infrared observations are crucial for studies of the structure of edge-on
galaxies. The all-sky 2MASS survey gives a good opportunity to enhance the
number of edge-on galaxies available for studies in the near-infrared bands.
Whereas faint parts of the galaxies are not seen in the 2MASS images, the 
thin disks of the galaxies are obtained with sufficient signal-to-noise
(S/N).

We refer to the Revised Flat Galaxies Catalog (RFGC; it contains extragalactic
objects with major-to-minor axis ratio a/b $>$ 7) as to a source of edge-on
galaxies. More than 200 RFGC galaxies with major axis size more than
1' were selected. This size was estimated directly from the 2MASS
images at the level S/N $\sim$ 3 and is significantly less that commonly 
used diameter D$_{25}$. Of them, there are 140 galaxies taken in all three 
2MASS bands: H, J, and K$_s$.
We applied the technique described by \citet{bm02} to assess the structural
parameters of edge-on disks: the radial (h) and vertical (z$_0$) scales, and
central face-on surface brightness S$_0$. This method implies an analysis of
photometric profiles drawn parallel to the major and minor axes of galaxy at
a one-pixel interval.  The h, z$_0$, and S$_0$ were obtained in the H, J,
and K$_s$ bands for selected 140 galaxies.

\section{Results and Discussion}

The dust layer in spiral galaxies is thinner than the stellar
disk \citep[see][]{xi99}. Since dust extinction is more significant in the J,
and less important in the K$_s$, we expect to obtain smaller scaleheights
for the same galaxies in the K$_s$ and larger in the J. One of the program
galaxy, RFGC 2946, is shown in Figure 1 in the J (top panel), H (middle),
and K$_s$ (bottom) bands.
In Figure 2 (upper panel) one can see the scaleheights in different 2MASS
bands for all considered edge-on galaxies. The scales in all bands are
normalized by the scaleheight in the J. Histograms in lower panel of Figure
2 show the same normalized slaleheights in the H (dotted line) and K$_s$
(dashed line). The mean relative ratios of the scaleheights are
1.00:0.91:0.86 in J:H:K$_s$ bands, respectively.

The different values of scaleheights z$_0$ for different bands enabled us to
infer the structural parameters of our edge-on galaxies under assumption of
zero extinction, i.e. if they had no dust.
Here we assume that the ratio of scaleheights is a linear function of the
internal extinction coefficient in given photometric band. It can be shown
that this approximation should work well for small optical depths, i.e. for
the case of NIR photometry). Hence, one can figure out the extinction-free
z$_0$ for the case of zero extinction coefficient. The resulting value of
the extinction-free scaleheights is, on average, 0.76 in units of z$_0$(J).
Hence, the K$_s$ scaleheight of thin stellar disks assessed in the K$_s$
band overestimates the original scaleleheight by, on average, only 13 \%.
Because of the large extinction, the scaleheights in the optical photometric
bands much more significantly overestimate the real vertical scales of the
stellar disks.

As it was shown in \citet{b00,bm02,kregel05}, the stellar disks with dimmer
central surface brightness (S$_0$) tend to have smaller ratios of scales
(z$_0$/h). The dust extinction may ne responsible for at least a part of
this relation.  Now we can check the relation z$_0$/h versus S$_0$ using the
extinction-free scaleheights. The central surface brightness in the K$_s$ band
is considered as S$_0$. Since the scalelength for our galaxies does not 
indicate a clear systematic variation between the J, H, and K$_s$ bands, 
we consider the K$_s$ scalelength here as the "h". Figure 3 suggests that 
correlation between the vertical to radial scales ratio (z$_0$/h) and 
disks' S$_0$ is not due to dust, but has a physical reason.

\acknowledgments
This project was partially supported by grant RFBR 04-02-16518.
This publication makes use of data products from the Two Micron All Sky
Survey, which is a joint project of the University of Massachusetts and the
Infrared Processing and Analysis Center/California Institute of Technology,
funded by the NASA and the NSF.


\begin{figure}
\plotone{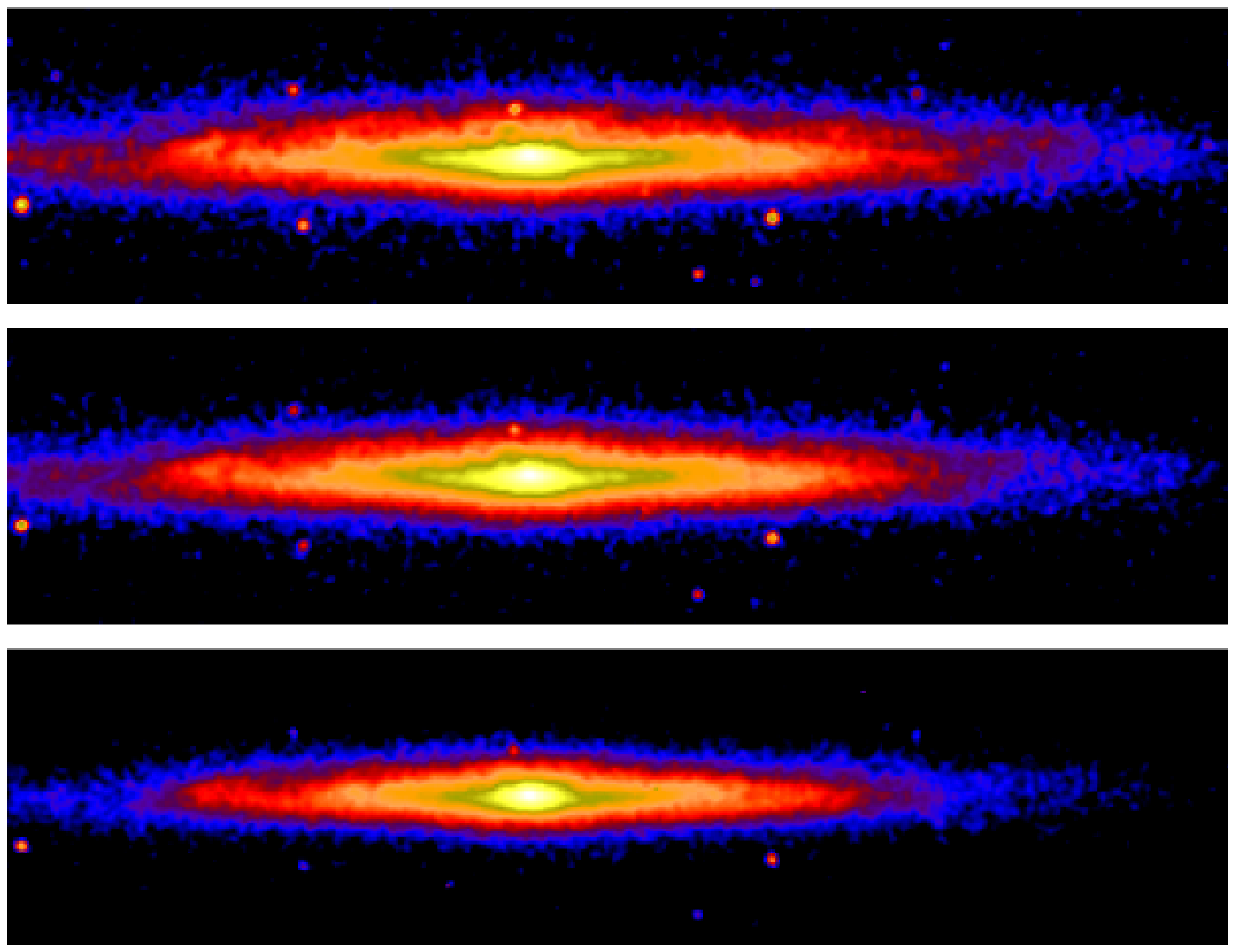}
\caption{Edge-on galaxy RFGC 2946 in the J (top), H (middle),
and K$_s$ (bottom) bands.
\label{fig1}}
\end{figure} 

\begin{figure}
\plotone{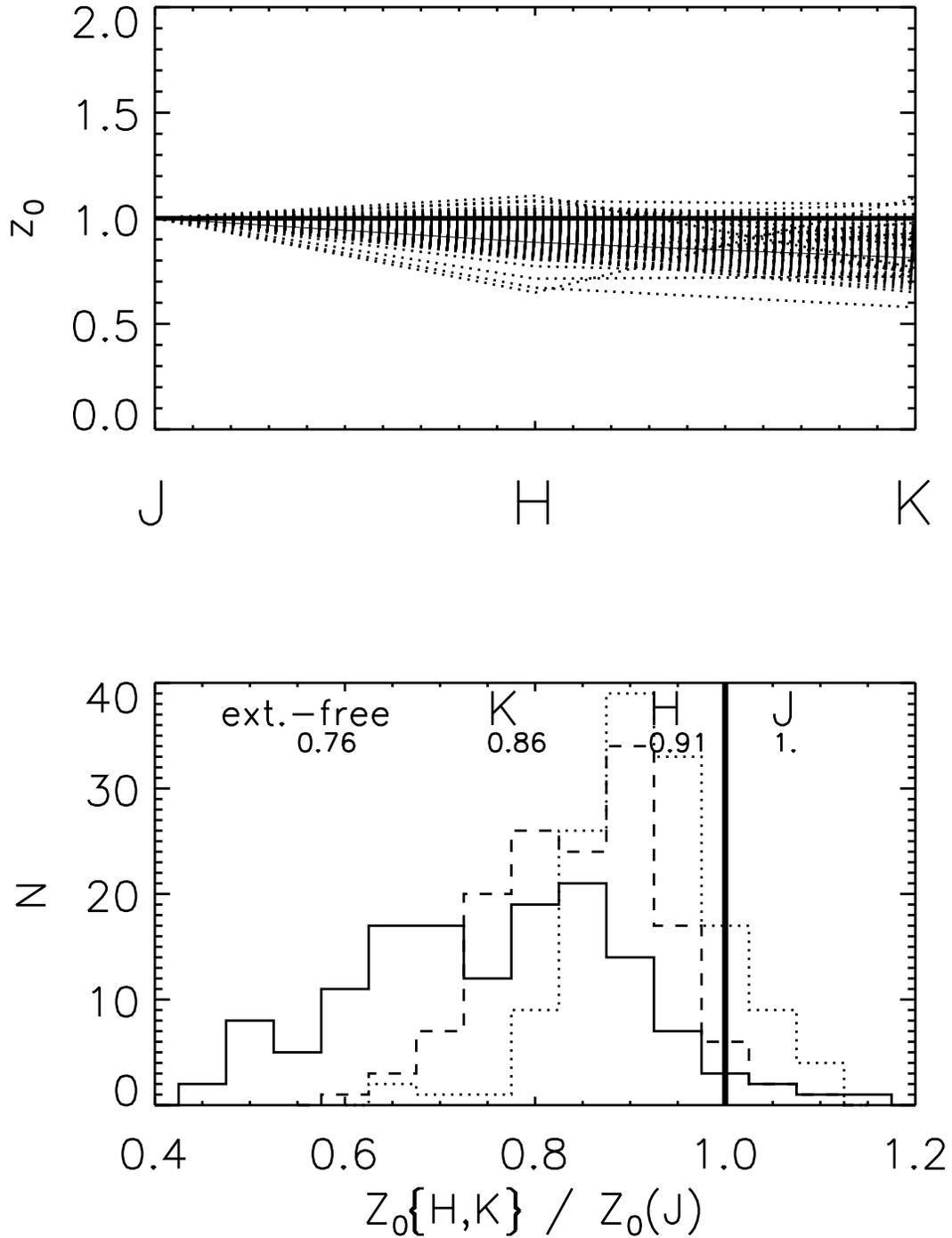}
\caption{Scaleheights z$_0$ in 2MASS bands J, H, and K$_s$ for all considered 
edge-on galaxies (top). All scales in all bands are normalized by the 
scaleheights in the J. Histograms in the bottom panel show the same 
normalized slaleheights in the H (dotted line), K$_s$ (dashed line), and
extinction-free (solid).
\label{fig2}}  
\end{figure}   

\begin{figure}
\plotone{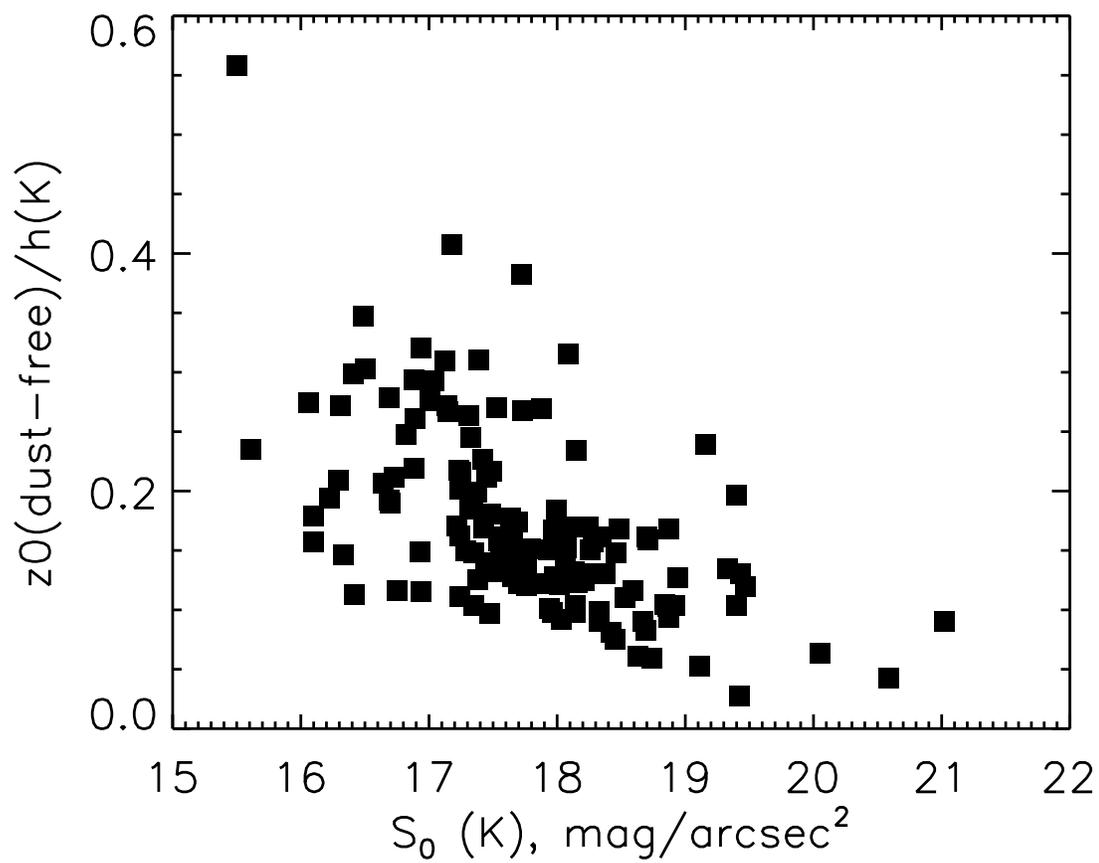}
\caption{Relation between the vertical to radial scales ratio (z$_0$/h) 
and the deprojected disks' central surface brightness S$_0$.     
\label{fig3}}  
\end{figure}

\end{document}